\documentclass[11pt]{article}
\usepackage{slashed}
\usepackage{amsmath}
\usepackage{textcomp}
\usepackage{amssymb}
\usepackage{amsfonts}
\usepackage{indentfirst}
\usepackage{xcolor}
\usepackage[colorlinks,citecolor=black,urlcolor=black,linkcolor=black]{hyperref}
\usepackage[square,comma,sort&compress,numbers]{natbib}
\usepackage[top=2cm,bottom=2cm,left=3cm,right=2cm]{geometry}
\usepackage{subcaption}

\newcommand{\be}{\begin{equation}}
\newcommand{\ee}{\end{equation}}
\newcommand{\bea}{\begin{eqnarray}}
\newcommand{\eea}{\end{eqnarray}}

\newcommand{\nn}{\nonumber\\}

\usepackage{graphicx,graphics}
\graphicspath{{figures/}}
\usepackage{calc}
\usepackage[normalem]{ulem}

\def\d{\mathrm{d}}

\def\C{\mathcal{C}}

\def\P{\mathcal{P}}
\def\T{\mathcal{T}}

\def\epsilon{\varepsilon}
\newcommand\blfootnote[1]{%
  \begingroup
  \renewcommand\thefootnote{}\footnote{#1}%
  \addtocounter{footnote}{-1}%
  \endgroup
}

\begin{document}

\thispagestyle{empty}

\begin{flushright}
{
\small
KCL-PH-TH/2022-53\\
}
\end{flushright}

\vspace{0.5cm}

\begin{center}
\Large\bf\boldmath
Wilsonian approach to the interaction $\phi^2(i\phi)^\varepsilon$
\unboldmath
\end{center}

\vspace{-0.2cm}

\begin{center}
Wen-Yuan Ai,*\blfootnote{*~wenyuan.ai@kcl.ac.uk} Jean Alexandre$^\dag$\blfootnote{$^\dag$~jean.alexandre@kcl.ac.uk} and Sarben Sarkar$^\ddag$\blfootnote{$^\ddag$~sarben.sarkar@kcl.ac.uk} \\

\vskip0.4cm

{\it Theoretical Particle Physics and Cosmology, King’s College London,\\ Strand, London WC2R 2LS, UK}\par

\vskip1.5cm
\end{center}

\begin{abstract}

We study the renormalisation of the non-Hermitian $\P\T$-symmetric scalar field theory with the interaction $\phi^2(i\phi)^\varepsilon$ using the Wilsonian approach and without any expansion in $\varepsilon$. Specifically, we solve the Wetterich equation in the local potential approximation, both in the ultraviolet regime and with the loop expansion. We calculate the scale-dependent effective potential and its infrared limit. The theory is found to be renormalisable at the one-loop level  only for integer values of $\varepsilon$, a result which is not yet established within the $\varepsilon$-expansion. Particular attention is therefore paid to the two interesting cases $\varepsilon=1,2$, and the one-loop beta functions for the coupling associated with the interaction $i\phi^3$ and $-\phi^4$ are computed. It is found that the $-\phi^4$ theory has asymptotic freedom in four-dimensional spacetime. Some general properties for the Euclidean partition function and $n$-point functions are also derived.  

\end{abstract}

\newpage

\hrule
\tableofcontents
\vskip.5cm
\hrule

\section{Introduction}

A real energy spectrum does not necessarily require the Hamiltonian to be Hermitian. Indeed, it was found by Bender and Boettcher~\cite{Bender:1998ke} that there are a large variety of Hamiltonians with $\P\T$ symmetry that can assure a real energy spectrum. Since then, quantum mechanics extended outside the Hermitian regime~\cite{Bender:2002vv} has become an active subject~\cite{Bender:2007nj,christodoulides2018parity,Bender:2019cwm}. In particular, non-Hermitian $\P\T$-symmetric Hamiltonians have found novel applications in condensed matter physics. See Refs.~\cite{longhi2018parity,el2018non,Ashida:2020dkc} for reviews. In recent years, $\P\T$-symmetric Hamiltonians have also witnessed increasing interest in high-energy physics. The consideration of non-Hermitian Hamiltonians may provide new mechanisms for neutrino masses and oscillations~\cite{Jones-Smith:2009qeu,Ohlsson:2015xsa,Alexandre:2015kra}, dark matter~\cite{Rodionov:2017dqt}, Higgs decay~\cite{Korchin:2016rsf}, and the confinement/deconfinement phase transition in QCD~\cite{Raval:2018kqg}. The generalisation of spontaneous symmetry breaking and the Goldstone theorem to non-Hermitian field theories has been carried out in Refs.~\cite{Alexandre:2017foi,Alexandre:2018uol,Mannheim:2018dur,Alexandre:2018xyy,Fring:2019hue,Fring:2019xgw,Alexandre:2019jdb,Fring:2020bvr}. Non-Hermitian Yukawa interactions with interesting phenomenological applications have been considered in Refs.~\cite{Alexandre:2020bet,Alexandre:2020tba,Mavromatos:2020hfy,Mavromatos:2021hpe}. Studies of the second quantisation and inner product in Fock space are given for a 
$\P\T$-symmetric scalar model in Ref.~\cite{Alexandre:2020gah} and for a $\P\T$-symmetric fermionic model in Ref.~\cite{Alexandre:2022uns}. For some other studies, see e.g. Refs.~\cite{Beygi:2019qab,Alexandre:2020wki,Felski:2020vrm,Fring:2020xpi,Chernodub:2020cew,Fring:2021zci,Felski:2021bdg,Felski:2021evi,Khunjua:2021fus,Mavromatos:2022heh,Felski:2022dsx,Naon:2022xvl,Gubaeva:2022feb}.

Non-Hermitian $\P\T$-symmeric theories are mostly well understood in quantum mechanics, especially for the well-studied model
\begin{align}
\label{eq:H}
    H=p^2+\frac{1}{2}\mu^2 x^2+\frac{1}{2} x^2(i x)^\varepsilon~.
\end{align}
For $\varepsilon\ge0$ the energy spectrum of the Hamiltonian was found to be real numerically~\cite{Bender:1998ke}. For the massless case spectral reality was proved for $\varepsilon>0$ by Dorey {\it et al.} using the methods of integrable systems~\cite{Dorey:2001uw}. The particular massless $\varepsilon=2$ case can be mapped to a Hermitian Hamiltonian with the same spectrum~\cite{Jones:2006qs,Bender:2006wt}. These results are based on the Schr\"{o}dinger equation directly.   

For higher spacetime-dimensional quantum field theories, the Schr\"{o}dinger equation is of functional type and very little information can be extracted from it. Therefore, alternative methods must be sought. A particularly useful tool is the path-integral formulation of quantum theories. Some earlier studies on $\P\T$-symmetric theories using the path integral are found in Refs.~\cite{Bender:2018pbv,Bender:2021fxa}. Recently,~in Ref.~\cite{Ai:2022csx} based on the Euclidean path integral, a new perspective that relates a non-Hermitian $\P\T$-symmetric theory  to a Hermitian theory via analytic continuation is given. In this way, conclusions for non-Hermitian theories could be drawn from the corresponding Hermitian theories. The relation proposed in Ref.~\cite{Ai:2022csx} assures that the Hamiltonian of form in Eq.~\eqref{eq:H} for $\varepsilon=2$ has a real spectrum even for spacetime dimension greater than one.

Compared with quantum-mechanical models, a new feature of (continuum) quantum field theory models is the presence of divergences due to the infinite number of degrees of freedom. Therefore, when extending non-Hermitian $\P\T$-symmetric quantum-mechanical models to field theory, one has to check that the theory is renormalisable, which is the topic of the present article. 
The theory we consider is the analogue of Eq.~\eqref{eq:H}, i.e., a scalar field theory with the bare potential at some cut-off scale $\Lambda$,
\be
\label{Ubare}
U_\Lambda(\phi)=\frac{1}{2}\mu^2\phi^2+\frac{1}{2}m^2\phi^2(i\phi/\phi_0)^\varepsilon~,
\ee
where $\phi_0$ is some scale, $\varepsilon\geq 0$, and $\mu^2>0$. 
Recently, there have been attempts to renormalise this theory~\cite{Bender:2018pbv,Felski:2021evi,Branchina:2021czr}. 
These attempts are based on an expansion in $\varepsilon$~\cite{Bender:1987dn,Bender:1988rq}. 
Although very interesting, these works have independently come to a puzzling conclusion that, at least within the $\varepsilon$-expansion, the theory  seems to be trivial ( at low orders in the expansion) for spacetime dimension $d=2$~\cite{Felski:2021evi} and $d\geq 2$~\cite{Branchina:2021czr}. Alternatively, these findings may indicate that the $\varepsilon$-expansion may not be valid for a systematic study of the renormalisation of the theory. The present work aims to study the renormalisation without an expansion in $\varepsilon$. 

We base our study on the Wilsonian framework, which in principle allows a non-perturbative description of quantum fluctuations, via Exact Renormalisation Group (ERG) equations.
The first ERG equation was derived by Wegner and Houghton~\cite{Wegner:1972ih}. Although very intuitive, this construction allows the evolution of the non-derivative part of the 
Wilsonian effective action only. 
An alternative ERG equation was proposed by Polchinski~\cite{Polchinski:1983gv} who introduced a smooth cut-off function for Fourier modes, and therefore the derivation of flows for the whole running Wilsonian action is allowed. 
Wetterich proposed a third approach~\cite{Wetterich:2001kra}, which elegantly combines the concept of a smooth cut-off and the one-particle irreducible (1PI) technique, through the ``average effective action". We will focus here on this third version of ERG equations. These equations always require
some approximation to be solved, and in this article we focus on either the ultraviolet (UV) regime or the one-loop regime under the local potential approximation.

The outline of the paper is as follows. In the next section, we describe some generic properties of the Euclidean path integral for this model and construct the 1PI effective action. We also summarise the main features of the Wetterich average effective action. In Section 3, we study the UV regime of the ERG equation, which can be mapped to a diffusion equation, and thus provides an intuitive understanding of how quantum fluctuations build up along the Wilsonian flow toward the infrared (IR). We show that the solution is analytical in the field for integer values of $\varepsilon$ only, which indicates potential consistency problems for non-integer $\varepsilon$. We then focus on $\varepsilon=1$ and discuss the beta function for the corresponding cubic coupling which, as expected, has the opposite sign compared to the Hermitian cubic interaction. Section 4 focuses on the one-loop Wilsonian flow which, by construction, recovers the one-loop 1PI effective potential in the deep IR limit. The latter potential contains new interactions for non-integer $\varepsilon$, with diverging coefficients, which is not consistent with renormalisability.
Only for integer $\varepsilon$ can one absorb divergences in a redefinition of bare parameters, and we give the explicit one-loop renormalisation for 
$\varepsilon=1$. Section 5 is devoted to the special case $\varepsilon=2$ in which a deformation of the integration contour is necessary to define a convergent path integral. We explain that the $\P\T$ symmetry is respected if the deformed contour is invariant under the $\P\T$-reflection. The construction of the 1PI effective action is however not modified and the results derived for a generic $\varepsilon$ can then be used for $\varepsilon=2$. We confirm that the interaction $-\phi^4$ is asymptotically free, unlike in the usual $+\phi^4$ theory. We conclude in Section~\ref{sec:con}.

\section{Properties of the quantum theory}

\subsection{Path integral convergence}

We consider the Euclidean partition function
\be
\label{eq:theory}
Z=\int\mathcal{D}[\phi]\, \exp\left(-\int \d^dx\left[\frac{1}{2}\partial_\mu\phi\partial^\mu\phi+U_\Lambda(\phi)\right]\right)~,
\ee
where $U_{\Lambda}(\phi)$ is given in Eq.~\eqref{Ubare}.
In the Hermitian case the scalar $\phi$ is supposed to be real. But in the non-Hermitian theory~\eqref{eq:theory} 
the path integral is convergent (with a UV cut-off) only for $\varepsilon$ restricted to specific intervals.\footnote{In \cite{Bender:2018pbv,Felski:2021evi,Branchina:2021czr} the path integral is defined through a formal expansion in the parameter $\varepsilon$. Although formally each term in the $\varepsilon$ expansion is calculated in terms of a convergent path integral, the properties and convergence of the series in $\varepsilon$ are unknown. Truncating to low order in $\varepsilon$ is in general an uncontrolled approximation.}
To see this, consider a real $\phi$ and write the self-interaction term as
\be
m^2\frac{|\phi|^{2+\varepsilon}}{\phi_0^\varepsilon}\Big(\cos(\pi\varepsilon/2+\theta\varepsilon)+i\sin(\pi\varepsilon/2+\theta\varepsilon)\Big)~,
\ee
where $\theta=0$ if $\phi\geq 0$ and $\theta=\pi$ if $\phi<0$. To make sure the real part is not negative for $|\phi|\rightarrow\infty$, we need to require 
\be\label{eq:requirement}
-1+4N\le\varepsilon\le 1+4N~~~~\mbox{and}~~~~\frac{1}{3}(-1+4N')\le\varepsilon\le\frac{1}{3}(1+4N')~,
\ee
where $N$ and $N'$ are integers. If we impose $\varepsilon$ to be positive, the allowed values/intervals are $\varepsilon\in[0,1/3]$; $\varepsilon=1$;
$\varepsilon\in[11/3,13/3]$; etc. For real $\phi$, the Lagrangian is invariant under the combined $\P\T$ operation where 
\bea
\P&:&~~~~ \phi(t,\vec{x})~\to~ -\phi(t,-\vec{x})~,\\
\T&:&~~~~ \phi(t,\vec{x})~\to~~~~ \phi(-t,\vec{x})~~~~\mbox{and}~~~~i~\to~-i~.\nonumber
\eea

For $\varepsilon$ that does not fall into the regions given in Eq.~\eqref{eq:requirement}, e.g. $\varepsilon=2$, $\phi$ necessarily takes values in the complex domain $\mathbb{C}$ to ensure the path integral to be convergent. In such a case, one may obtain the theory by analytically continuing $\varepsilon$ from the regions given in Eq.~\eqref{eq:requirement} to the interested value. If one studies the theory directly with the path integral, one in principle should apply the Picard-Lefschetz theory~\cite{pham1983vanishing,berry1991hyperasymptotics,Witten:2010cx,Witten:2010zr,Ai:2019fri} for the path integral with first complexifying the field configurations $\{\phi(x)\}\rightarrow\{\Phi(x)\}$ (where $\Phi$ takes values in $\mathbb{C}$) and then finding a middle-dimensional contour $\C_{\P\T}$ in the path integral, with the requirements that the $\P\T$ symmetry is still respected and the path integral is convergent.  Therefore, one ends up with
\be
\label{eq:path-integral}
Z=\int_{\C_{\P\T}}{\cal D}[\Phi]\exp\left(-\frac{1}{2}\int \d^d x\,\left[\partial_\mu\Phi\partial^\mu\Phi+\mu^2\Phi^2+m^2\Phi^2(i\Phi/\phi_0)^\varepsilon\right]\right)~.
\ee
Since $\Phi(x)$ takes in general complex values now, the Lagrangian in~\eqref{eq:path-integral} is not $\P\T$-symmetric anymore because under $\T$, one has to take in addition the complex conjugate of $\Phi$. However, one may implement the $\P\T$ symmetry for the path integral as a whole. Performing the $\P\T$ operation for the path integral, one obtains 
\be
\widetilde{Z}=\int_{\widetilde{\C}_{\P\T}}{\cal D}[\widetilde{\Phi}]\exp\left(-\frac{1}{2}\int \d^d x\,\left[\partial_\mu\widetilde{\Phi}\partial^\mu\widetilde{\Phi}+\mu^2\widetilde{\Phi}^2+m^2\widetilde{\Phi}^2(-i\widetilde{\Phi}/\phi_0)^\varepsilon\right]\right)~,
\ee
where $\widetilde{\Phi}=-\Phi^*$ and $\widetilde{\C}_{\P\T}=\{\widetilde{\Phi}(x): -\widetilde{\Phi}^*(x)\in \C_{\P\T}\}$. 
For $\varepsilon=2N$, the path integral is invariant under $\P\T$ if
\be
\label{eq:cond-contour}
\widetilde{\C}_{\P\T}=\C_{\P\T}~.
\ee
For the zero-dimensional case, such a contour satisfy the so-called left-right symmetry in the complex plane. For other values of $\varepsilon$, a more delicate analysis of the contour is required. Note that doing the above procedure one is not adding more degrees of freedom to the theory because the middle-dimensional contour $\C_{\P\T}$ has the same ``dimension'' as that of the original real configuration space.\footnote{Rigorously speaking, the dimension of the contour $\C_{\P\T}$ is infinity.}

In this article, we study the functional renormalisation of the theory~\eqref{eq:theory}. 
We are in particular interested in the region $\varepsilon\in [0,2]$. 
We carry out the analysis first for $\varepsilon\in [0,1/3]$ and $\varepsilon=1$ in which $\phi$ is kept real and then analytically continue the results to other values out of these regions. As we shall see below, one-loop divergences can be absorbed by counter-terms for integer values of $\varepsilon$ only.

\subsection{One-particle-irreducible effective action}

We assume here either $\varepsilon\in [0,1/3]$ or $\varepsilon=1$, in which case the path integration is done over real $\phi$ configurations.
Given a (real) source $J$, we define the Euclidean partition function in such a way that the source term is invariant under $\P\T$ symmetry
\be
\label{eq:sourced-Z}
Z[J]=\int {\cal D}[\phi]\exp\left(-\frac{1}{2}\int \d^d x~\left[\partial_\mu\phi\partial^\mu\phi+\mu^2\phi^2+m^2\phi^2(i\phi/\phi_0)^\varepsilon\right]
-i\int \d^d x~ J\phi\right)~.
\ee
The one-point function is defined from the connected generating functional $W[J]=-\ln Z[J]$ as
\be\label{one-point}
\varphi\equiv \left<\phi\right> =\frac{\delta W}{i\delta J}=-\frac{1}{Z}\frac{\delta Z}{i\delta J}~,
\ee
where
\be
\left<\cdots\right>\equiv\frac{1}{Z}
\int {\cal D}[\phi](\cdots)\exp\left(-\frac{1}{2}\int \d^d x~\left[\partial_\mu\phi\partial^\mu\phi+\mu^2\phi^2+m^2\phi^2(i\phi/\phi_0)^\varepsilon\right]
-i\int \d^d x~J\phi\right)~.
\ee
Taking the complex conjugate of $Z[J]$, one obtains
\be
(Z[J])^*=\int {\cal D}[\phi]\exp\left(-\frac{1}{2}\int \d^d x\,\left[\partial_\mu\phi\partial^\mu\phi+\mu^2\phi^2+m^2\phi^2(-i\phi/\phi_0)^\varepsilon\right]
+i\int \d^d x~J\phi\right)~,
\ee
and the change of variable, $\phi\rightarrow -\phi$ leads to 
\be
(Z[J])^*=Z[J]~.
\ee
Similarly, we have
\be
(\varphi[J])^*=-\varphi[J]~,
\ee
such that the one-point function $\varphi[J]$ is purely imaginary, as long as the source $J$ is real. Actually, one can extend this argument to arbitrary $n$-point functions. It can be seen that any $2N$-point correlation function (with $N$ being a non-negative integer) is real and any $2N+1$-point correlation function is imaginary. 

We also have the usual relation 
\be
\frac{\delta^2 W}{i\delta J(x) i\delta J(y)}=\varphi(x)\varphi(y)-\left<\phi(x)\phi(y)\right>~,
\ee
which expresses the second functional derivative of $W$ in terms of the variance of quantum fluctuations. 
One then inverts the relation $\varphi[J]\to J[\varphi]$ in order to define the Legendre transform 
\be
\Gamma[\varphi]=W[J[\varphi]]-i\int \d^4x~\varphi J[\varphi]~,
\ee
which represents the 1PI effective action, with functional derivatives
\bea
\frac{\delta\Gamma[\varphi]}{\delta\varphi(x)}&=&-i J(x)~,\\
\frac{\delta^2 \Gamma[\varphi]}{\delta\varphi(x)\delta\varphi(y)}&=&
-\left(\frac{\delta^2 W}{i\delta J(x) i\delta J(y)}\right)^{-1}~.\nonumber
\eea
From the above equations we finally obtain
\be\label{delta2Gamma}
\frac{\delta^2 \Gamma[\varphi]}{\delta\varphi(x)\delta\varphi(y)}=\Big(\left<\phi(x)\phi(y)\right>-\varphi(x)\varphi(y)\Big)^{-1}~.
\ee

In summary, we find that the $\P\T$-symmetric theory~\eqref{eq:sourced-Z} has real $Z[J]$ as well as real 1PI effective action $\Gamma[\varphi]$. 
However, although $\phi$ is real, its one-point function $\varphi$ is purely imaginary.
Note that the physically relevant coupling constants are obtained from the derivatives $\delta^n\Gamma/\delta\varphi^n$ at $\varphi=0$, 
and do not depend on $\varphi$ being purely imaginary.

\subsection{Exact Wilsonian renormalisation}

The Wilsonian evolution of the Wetterich running action defined at some scale $k$ is derived in the framework of the 1PI quantisation, 
where a cut-off function is added to the bare action, in order to ``freeze" infrared modes with momentum $|p|\lesssim k$ in the path integration~\cite{Wetterich:2001kra}.
This is achieved through the term
\be
\frac{1}{2}\int\frac{\d^d p}{(2\pi)^d}\tilde\phi(p)R_k(p)\tilde\phi(-p)~,
\ee
where $\tilde\phi(p)$ is the Fourier transform of the field $\phi(x)$. The function $R_k(p)$ is not unique, but it vanishes for $k\to0$,
such that this limit reproduces the usual 1PI quantisation. In the usual Hermitian context,
the corresponding 1PI ``average effective action" $\Gamma_k$ satisfies the exact functional renormalisation equation 
\be\label{originalERG}
\partial_k\Gamma_k
=\frac{1}{2}\int\frac{\d^d p}{(2\pi)^d}\partial_kR_k(p)\left(\frac{\delta^2\Gamma_k}{\delta\tilde\varphi(p)\delta\tilde\varphi(-p)} +R_k(p)\right)^{-1}~,
\ee
where $\tilde\varphi$ is the Fourier transform of the one-point function. 
We also note that, by construction, the IR limit $k\to0$ of $\Gamma_k$ reproduces the one-particle
irreducible (1PI) effective action~\cite{Jackiw:1974cv}, which is independent of the blocking procedure in Fourier space.

In our situation, 
assuming either $\varepsilon\in [0,1/3]$ or $\varepsilon=1$, the change $J\to iJ$ doesn't modify the derivation of the 
equation (\ref{originalERG}), which therefore remains the same, but where $\varphi(x)$ is purely imaginary, such that
$\tilde\varphi(-p)=-\tilde\varphi^*(p)$.

We choose the Litim cut-off function~\cite{Litim:2001up}, and work in the local potential approximation, where the evolution of derivative terms are neglected and the running effective action takes the form
\be\label{localapprox}
\Gamma_k[\varphi]\equiv\int \d^dx\left[\frac{1}{2}\partial_\mu\varphi\partial^\mu\varphi+U_k(\varphi)\right]~.
\ee
The resulting Exact Renormalisation Equation (ERG) is
\be\label{localERG}
\partial_k U_k(\varphi)=\frac{\alpha_d\, k^{d+1}}{k^2+U_k''(\varphi)}~,
\ee
where 
\be
\alpha_d\equiv\frac{\hbar~\Omega_d}{d(2\pi)^d}~,
\ee
and $\Omega_d\equiv 2\pi^{d/2}/\Gamma[d/2]$ is the solid angle in dimension $d$.
In Eq.~\eqref{localERG}, $U_k$ is the running potential for $0\le k\le\Lambda$ and a prime denotes a derivative with respect to $\varphi$. 
This equation ignores the renormalisation of the derivative terms in the running action, but it provides a resummation of all quantum fluctuations in this approximation. This equation is a challenge to solve in the generic case but one can make the most of this resummation in some specific regimes, as we show in the next sections. 

For integer $\varepsilon$, one can read the beta-functions of a theory from Eq.~\eqref{localERG}. 
Assume a running interaction of the form $\lambda_k\varphi^n/n!$ where $n\ge3$ is an integer, then
\be\label{lambdak}
\lambda_k=\frac{\partial^n}{\partial\varphi^n}\Big( U_k(\varphi)\Big)_{\varphi=0}~.
\ee
The mass dimension of $\lambda_k$ is $[\lambda]=d-n(d/2-1)$, and one defines the dimensionless coupling by rescaling $\lambda_k$ with 
the appropriate power of $k$
\be
\tilde \lambda_k\equiv k^{-[\lambda]}\lambda_k~.
\ee
The corresponding beta-function is then
\be\label{beta}
\beta\equiv k\partial_k\tilde \lambda_k=-[\lambda]\tilde \lambda_k
+k^{-[\lambda]}\frac{\partial^n}{\partial\varphi^n} \Big(k\partial_k U_k(\varphi)\Big)_{\varphi=0}~,
\ee
where $\partial_k U_k(\varphi)$ is obtained from Eq.~\eqref{localERG}. The first term on the RHS of Eq.~\eqref{beta} corresponds to the trivial scaling law for $\lambda_k$, and the second term corresponds to the anomalous dimension, arising from quantum fluctuations.

\section{Ultraviolet regime}

In this section, we focus on the UV behaviour of the running potential. In the UV regime where $\Lambda^2\ge k^2 \gg |U_k''(\phi_0)|\sim \mu^2+m^2$, the ERG equation can then be written as
\be\label{ERGUV}
\partial_k U_k(\varphi)=\alpha_d k^{d-1}\sum_{n=0}^\infty (-1)^n k^{-2n}\Big(U_k''(\varphi)\Big)^n~.
\ee

\subsection{Diffusion}

If we introduce the notation 
\be
\hat U_k(\varphi)\equiv U_k(\varphi)-\frac{\alpha_d}{d}k^d~, 
\ee
and keep the dominant term $n=1$ on the RHS of Eq.~\eqref{ERGUV}, we obtain the diffusion equation\footnote{A related approach for $d=1$ was studied in \cite{Bender:2018xpo}.}
\be\label{diffusion}
\partial_\tau \hat U_\tau(\varphi)=\partial_\varphi^2 \hat U_\tau(\varphi)~,
\ee
where $\tau$ is defined as
\be
\frac{\d\tau}{\d k}=-\alpha_d k^{d-3}~.
\ee
In the present Wilsonian picture, the system gets dressed by quantum corrections as $k$ decreases from $\Lambda$, or with  
the above parameterisation, as $\tau$ increases from 0. Specifically,
\begin{align}
\label{eq:tau}
    \tau=
\begin{cases}
\alpha_2\ln \left(\frac{\Lambda}{k}\right)\,, \qquad\qquad\quad\;  {\rm d=2}\,, \\
\frac{\alpha_d}{d-2}\left(\Lambda^{d-2}-k^{d-2}\right)\,, \quad {\rm d\geq 3}\,.
\end{cases}  
\end{align}
A solution that is analytical at $\tau=0$ can be written formally as
\be
\hat U_\tau(\varphi)=\exp\Big(\tau\partial^2_\varphi\Big)\hat U_{\tau=0}(\varphi)~,
\ee
such that
\be\label{UkUV}
\hat U_k(\varphi)=\mu^2\tau+\frac{1}{2}\mu^2\varphi^2+\frac{1}{2}m^2\varphi^2(i\varphi/\phi_0)^\varepsilon
\sum_{n=0}^\infty\frac{1}{n!}\frac{f_n\,\tau^n}{\varphi^{2n}}~,
\ee
where $f_0=1$ and for $n\ge1$
\be
f_n=\prod_{p=0}^{n-1}(\varepsilon+2-2p)(\varepsilon+1-2p)~.
\ee
Hence for any non-integer $\varepsilon$, the sum (\ref{UkUV}) is infinite and is not analytical at $\varphi=0$. However, for any integer $\varepsilon$ the sum (\ref{UkUV}) is finite and analytical at $\varphi=0$. Also, one can check that the quadratic potential for $\varepsilon=0$ does not get quantum corrections:
\bea
\hat U_k(\varphi)&=&\mu^2\tau+\frac{1}{2}\mu^2\varphi^2+\frac{1}{2}m^2\varphi^2\left(1+\frac{2\tau}{\varphi^2}\right)\\
&=&\frac{1}{2}(\mu^2+m^2)\varphi^2+~\mbox{$\varphi$-independent terms}~,\nonumber
\eea
which is expected.

\subsection{Beta-function for the interaction \texorpdfstring{$i\phi^3$}{TEXT}}

For $\varepsilon=1$, the cubic coupling is defined as
\be
i\lambda_k\equiv \left(\frac{\partial^3U_k(\varphi)}{\partial\varphi^3}\right)_{\varphi=0}~,
\ee
and has the bare value $\lambda_\Lambda=3m^2/\phi_0$. The dimensionless coupling is $\tilde\lambda_k=k^{d/2-3}\lambda_k$ with the corresponding 
beta-function
\be\label{beta3}
\beta\equiv k\partial_k\tilde \lambda_k=(d/2-3)\tilde \lambda_k
-ik^{d/2-3}\frac{\partial^3}{\partial\varphi^3} \Big(k\partial_k U_k(\varphi)\Big)_{\varphi=0}~.
\ee
From the result (\ref{UkUV}) for $\varepsilon=1$, the UV running potential is
\bea\label{UVe=1}
\hat U_k(\varphi)
%&=&\mu^2\tau+\frac{1}{2}\mu^2\varphi^2+\frac{1}{2}m^2\varphi^2(i\varphi/\phi_0)\left(1+\frac{6\tau}{\varphi^2}\right)\\
=\frac{1}{2}\mu^2\varphi^2+\frac{1}{2}m^2\varphi^2(i\varphi/\phi_0)+\tau\Big(\mu^2+3m^2(i\varphi/\phi_0)\Big)~,\nonumber
\eea
where we note that the third derivative with respect to $\varphi$ does not depend on $k$, such that naively the beta-function vanishes.
This is due to the truncation in Eq.~\eqref{ERGUV}, 
and we need to reintroduce the resummation before calculating the beta-function. 
In terms of the original potential $U_k(\varphi)$, we obtain 
%from the result (\ref{UVe=1})
\bea
\label{eq:UV-Uk-phi3}
\partial_k U_k(\varphi)&=&\alpha_d k^{d-1}+\frac{d\tau}{dk}\Big(\mu^2+3m^2(i\varphi/\phi_0)\Big)\notag \\
&=&\alpha_d k^{d-1}-\alpha_d k^{d-3}\Big(\mu^2+3m^2(i\varphi/\phi_0)\Big)\nn
&\simeq&\frac{\alpha_d\, k^{d+1}}{k^2+\mu^2+3m^2(i\varphi/\phi_0)}~,
\eea
which reproduces the one-loop evolution equation (details are given in the next section). As a consequence, the UV beta function coincides with the 
one-loop beta-function provided one performs the above resummation. It is interesting to note that the UV regime contains the same information as
the one-loop result, which shows the equivalence between the small-coupling regime and the UV regime, where
quantum fluctuations are perturbative.

From Eq.~\eqref{eq:UV-Uk-phi3}, one then obtains
\be
-i\left(\frac{\partial^3[k\partial_kU_k(\varphi)]}{\partial\varphi^3}\right)_{\phi=0}=162\alpha_d\frac{m^6}{\phi_0^3}\frac{k^{d+2}}{(k^2+\mu^2)^4}~.
\ee
Substituting the above into Eq.~\eqref{beta3}, one finally arrives at the beta-function in the UV regime,
\be
\label{betaUV}
\beta^{\rm (UV)}=(d/2-3)\tilde \lambda_k+\frac{6\alpha_d(\tilde \lambda_\Lambda)^3}{(1+(\tilde\mu)^2)^4}~,
\ee
where $\tilde\mu=\mu/k$. It is interesting to look at the specific case $d=6$, where
one can replace the bare coupling by the dressed coupling on the RHS of Eq.~\eqref{betaUV}, 
since the difference is of order $\hbar^2$. The beta function in the UV regime $\tilde\mu \ll 1$ is then
\be
\beta^{\rm (UV)}\simeq6\alpha_6\lambda_k^3=\frac{\hbar\lambda_k^3}{64\pi^3}~~~~\mbox{for}~~d=6~,
\ee
and is positive, unlike the corresponding beta-function for the real cubic interaction.

\subsection{Corrections to the diffusion equation}

Coming back to Eq.~\eqref{ERGUV}, we can write 
\be
\partial_\tau\hat U_\tau(\varphi)=\sum_{n=0}^\infty (-1)^n k^{-2n}\Big(\hat U_\tau''(\varphi)\Big)^{n+1}~.
\ee
Through the relation given in Eq.~\eqref{eq:tau}, one can perform a systematic perturbative expansion of the solution in powers of $\Lambda^{-2}$. 
For example, for $d=4$, we have
\be
\tau=\frac{\alpha_4}{2}(\Lambda^2-k^2)~,
\ee
such that the first-order correction to the diffusion equation reads, in the UV regime $\tau\ll\Lambda^2$,
\be
\partial_\tau\hat U_\tau(\varphi)=\hat U_\tau''(\varphi)-\frac{1}{\Lambda^2}\Big(\hat U_\tau''(\varphi)\Big)^2+{\cal O}(1/\Lambda^4)~.
\ee
The solution can then be expanded as
\be
\hat U_\tau(\varphi)=\hat U_\tau^{(0)}(\varphi)+\frac{1}{\Lambda^2}\hat U_\tau^{(1)}(\varphi)+{\cal O}(1/\Lambda^4)~,
\ee
where $\hat U_\tau^{(0)}$ satisfies the diffusion equation, and $\hat U_\tau^{(1)}$ satisfies 
\be
\partial_\tau \hat U^{(1)}_\tau(\varphi)=\partial_\varphi^2 \hat U^{(1)}_\tau(\varphi)-\Big( \partial^2_\varphi \hat U^{(0)}_\tau(\varphi)\Big)^2~.
\ee
We are interested in a particular solution only, which satisfies the appropriate boundary conditions, 
because the homogeneous equation is the diffusion equation, with a solution proportional to $\hat U^{(0)}_{\tau}(\varphi)$.
The latter equation can in principle be solved, but in what follows we turn to the one-loop approximation, without assuming $k$ to be large.

\section{One-loop approximation}
\label{Oneloop}

In this section, we truncate the ERG equation (\ref{localERG}) to one loop.

\subsection{Coupling constant}

If we introduce the notation $\xi\equiv\varphi/\phi_0$, the bare potential can be written
\be
U_\Lambda(\xi)=\frac{1}{2}\mu^2\phi_0^2~\xi^2+\frac{1}{2}m^2\phi_0^2~\xi^2(i\xi)^\varepsilon~,
\ee
and because the ERG equation (\ref{localERG}) is not linear, it induces a phase change in the running potential, 
unlike what happens in the UV regime. The running potential can be parametrised as
\be
U_k(\xi)=\frac{1}{2}m^2\phi_0^2~V_k(\xi)~,
\ee
such that in terms of $V_t(\xi)$ the ERG equation (\ref{localERG}) reads
\be\label{dimensionlessERG}
\partial_t V_t(\xi)=\frac{g\,t^{d+1}}{t^2+\partial_\xi^2 V_t(\xi)}~,
\ee
where 
\be
g\equiv\frac{\alpha_d\,m^{d-2}}{2^{d/2-1}\phi_0^2}~~~~\mbox{and}~~~~t=\frac{\sqrt{2}k}{m}~.
\ee
One can see that the dimensionless parameter $g$ plays the role of a coupling constant, and the evolution in $t$ 
of the running potential is proportional to $g$.
As a consequence, if $g\ll 1$, the ERG equation can be expanded in this coupling constant 
\be
\partial_t V_t(\xi)=\frac{g\,t^{d+1}}{t^2+\partial_\xi^2 V_T(\xi)}+{\cal O}(g^2)~,
\ee
where $T=\sqrt2 \Lambda/m$ and
\be
\partial_\xi^2 V_T(\xi)=2\mu^2/m^2+(2+\varepsilon)(1+\varepsilon)(i\xi)^\varepsilon~.
\ee
The evolution equation for the one-loop running potential $V_t^{(1)}(\xi)$ is finally
\be\label{V1}
\partial_t V_t^{(1)}(\xi)=\frac{g\,t^{d+1}}{t^2+\partial_\xi^2 V_T(\xi)}~.
\ee
This equation can be solved for arbitrary $d$. But below we will focus on the cases with $d=4$ and $d=2$. The calculation can of course be easily generalised to other cases with different values of $d$, e.g. $d=3$.

\subsection{One-loop effective potential (\texorpdfstring{$d=4$}{TEXT})}

The integration of Eq.~\eqref{V1} is straightforward, and for $d=4$ it leads to
\bea
V_t^{(1)}(\xi)&=&V_T(\xi)+g\int_T^t \frac{u^5~du}{u^2+\partial_\xi^2 V_T(\xi)}\\
&=&V_T(\xi)+\frac{g}{2}(T^2-t^2)\partial_\xi^2 V_T(\xi)
-\frac{g}{2}\Big(\partial_\xi^2 V_T(\xi)\Big)^2\ln\left(\frac{T^2+\partial_\xi^2 V_T(\xi)}{t^2+\partial_\xi^2 V_T(\xi)}\right)~,\nonumber
\eea
where a field-independent term is disregarded. In terms of the original variables, we obtain then
\bea
\label{eq:Uk-one-loop-4d}
U_k^{(1)}(\varphi) &=& \frac{1}{2}\mu^2\varphi^2+\frac{1}{2}m^2\varphi^2(i\varphi/\phi_0)^\varepsilon\notag +\frac{\hbar}{64 \pi^2}(\Lambda^2-k^2)\left[ \mu^2+ m^2(2+\varepsilon)(1+\varepsilon)(i\varphi/\phi_0)^\varepsilon/2 \right]\\
&&\qquad\qquad\qquad\qquad-\frac{\hbar}{64\pi^2}\left[\mu^2+m^2(2+\varepsilon)(1+\varepsilon)(i\varphi/\phi_0)^\varepsilon/2\right]^2 \notag\\
&&\qquad\qquad\qquad\qquad\times\ln\left(\frac{2\Lambda^2+2\mu^2+m^2(2+\varepsilon)(1+\varepsilon)(i\varphi/\phi_0)^\varepsilon}
{2k^2+2\mu^2+m^2(2+\varepsilon)(1+\varepsilon)(i\varphi/\phi_0)^\varepsilon}\right)~.
\eea
Finally, the one-loop Wilsonian effective potential is obtained in the limit $k\to 0$. If we ignore terms vanishing in the limit $m/\Lambda\to0$, it reads
\bea
\label{Ueff1exp-4d}
U_{\rm eff}^{(1)}(\varphi)&=&\frac{1}{2}\mu^2\varphi^2+\frac{1}{2}m^2\varphi^2(i\varphi/\phi_0)^\varepsilon
+\frac{\hbar}{64\pi^2}\Lambda^2\left[\mu^2+m^2(2+\varepsilon)(1+\varepsilon)(i\varphi/\phi_0)^\varepsilon/2\right]\nn
&&\qquad\qquad\qquad-\frac{\hbar}{64\pi^2}\left[\mu^2+m^2(2+\varepsilon)(1+\varepsilon)(i\varphi/\phi_0)^\varepsilon/2\right]^2\nn
&&\qquad\qquad\qquad\times\left[\ln\left(\frac{\Lambda^2}{m^2}\right)
-\ln\left(\frac{\mu^2}{m^2}+\frac{1}{2}(2+\varepsilon)(1+\varepsilon)(i\varphi/\phi_0)^\varepsilon\right)\right]~,
\eea
and is identical to the one-loop 1PI effective potential. 

We can note an important property: for a generic non-integer $\varepsilon$, one-loop corrections generate the new interactions $(i\varphi/\phi_0)^\varepsilon$
and $(i\varphi/\phi_0)^{2\varepsilon}$, with quadratic and logarithmic divergent coefficients respectively. 
Hence one cannot define counter-terms without changing the dynamics of the problem; the theory is not renormalisable for a generic non-integer $\varepsilon$.

One-loop renormalisability can be achieved for $\varepsilon=1$ though, where
\bea
U_{\rm eff}^{(1)}(\varphi)&=&\frac{1}{2}\mu^2\varphi^2+\frac{1}{2}m^2\varphi^2(i\varphi/\phi_0)
+\frac{\hbar}{64\pi^2}\Lambda^2\left[\mu^2+3m^2 i\varphi/\phi_0\right]\nn
&&-\frac{\hbar}{64\pi^2}\left(\mu^2+3m^2i\varphi/\phi_0\right)^2\left[
\ln\left(\frac{\Lambda^2}{m^2}\right)-\ln\left(\frac{\mu^2}{m^2}+3i\varphi/\phi_0\right)\right]~,
\eea
and contains tadpole terms (linear in $\varphi$) which can be eliminated with counter-terms with no modification of the dynamics. 
The $\varphi$-independent  divergent terms have no physical effect and can be discarded also.
The logarithmically divergent quadratic term can be absorbed in a redefinition of $\mu^2$ through the renormalised mass squared
\be
\mu_{\rm R}^2=\mu^2+\frac{9\hbar\,m^4}{32\pi^2\phi_0^2}\ln\left(\frac{\Lambda^2}{m^2}\right)~,
\ee
and the remaining terms are finite. The renormalised one-loop effective potential is then
\be
U_{\rm R}^{(1)}(\varphi)=\frac{1}{2}\mu_{\rm R}^2\varphi^2+\frac{1}{2}m^2\varphi^2(i\varphi/\phi_0)+
\frac{\hbar}{64\pi^2}\left(\mu_{\rm R}^2+3m^2i\varphi/\phi_0\right)^2\ln(\mu_{\rm R}^2/m^2+3i\varphi/\phi_0)~.
\ee
Note that one cannot consistently set $\mu^2=0$, since mass term corrections are generated for $\varepsilon=1$, 
with divergences that have to be absorbed in the bare mass term.

\subsection{One-loop effective potential (\texorpdfstring{$d=2$}{TEXT})}

For $d=2$ the integration of Eq.~\eqref{V1} leads to, up to a $\varphi$-independent term,
\be
V_t^{(1)}(\xi)=V_T(\xi)+\frac{g}{2}~\partial_\xi^2 V_T(\xi)~\ln\left(\frac{T^2+\partial_\xi^2 V_T(\xi)}{t^2+\partial_\xi^2 V_T(\xi)}\right)~,
\ee
which gives
\bea
\label{eq:Uk-one-loop-2d}
U_k^{(1)}(\varphi) &=& \frac{1}{2}\mu^2\varphi^2+\frac{1}{2}m^2\varphi^2(i\varphi/\phi_0)^\varepsilon\notag +\frac{\hbar}{8\pi}\left[\mu^2+m^2(2+\varepsilon)(1+\varepsilon)(i\varphi/\phi_0)^\varepsilon/2\right] \notag\\
&&\qquad\qquad\qquad\qquad\qquad\times\ln\left(\frac{2\Lambda^2+2\mu^2+m^2(2+\varepsilon)(1+\varepsilon)(i\varphi/\phi_0)^\varepsilon}
{2k^2+2\mu^2+m^2(2+\varepsilon)(1+\varepsilon)(i\varphi/\phi_0)^\varepsilon}\right)~.
\eea
Taking $k\rightarrow 0$ and ignoring terms vanishing in the limit $\Lambda\to\infty$ leads to
\bea
\label{eq:Ueff-2d}
U_{\rm eff}^{(1)}(\varphi)&=&\frac{1}{2}\mu^2\varphi^2+\frac{1}{2}m^2\varphi^2(i\varphi/\phi_0)^\varepsilon\\
&&+\frac{\hbar}{8\pi}\Big(\mu^2+m^2(2+\varepsilon)(1+\varepsilon)(i\varphi/\phi_0)^\varepsilon/2\Big)\nn
&&\qquad\times\left[\ln\left(\frac{\Lambda^2}{m^2}\right)-\ln\left(\frac{\mu^2}{m^2}+\frac{1}{2}(2+\varepsilon)(1+\varepsilon)(i\varphi/\phi_0)^\varepsilon\right)\right]~.\nonumber
\eea
As in the situation where $d=4$, this one-loop potential is renormalisable for $\varepsilon=1$ only, since only a tadpole appears to be divergent, 
which can be removed with a counter-term. The divergent $\varphi$-independent term is not physical and can be discarded. 
The resulting renormalised effective potential has new interactions, which are however all finite
\bea
U_{\rm R}^{(1)}(\varphi)&=&\frac{1}{2}\mu^2\varphi^2+\frac{1}{2}m^2\varphi^2(i\varphi/\phi_0)-\frac{\hbar}{8\pi}\Big(\mu^2+3m^2(i\varphi/\phi_0)\Big)
\ln\left(\frac{\mu^2}{m^2}+3(i\varphi/\phi_0)\right)~.
\eea

\section{Analytical continuation for \texorpdfstring{$\varepsilon=2$}{TEXT}}

In this section, we extend the analysis to $\varepsilon=2$. In this case, the integral~\eqref{eq:theory} is not convergent for real $\phi$ and therefore one needs to consider a contour different from the space of real configurations as discussed in Eq.~\eqref{eq:path-integral}. In order for the path integral to respect the $\P\T$ symmetry, it is sufficient to enforce the condition~\eqref{eq:cond-contour} on the contour $\C_{\P\T}$. Note that although seemly the potential is unbounded from below for $\epsilon=2$, the theory in the $\P\T$-symmetric framework actually has a stable ground state at least for $d=1$. The conjectural relation proposed in Ref.~\cite{Ai:2022csx} indicates that this is also the case in higher-dimensional spacetime.

\subsection{One-particle-irreducible effective action}

Introducing a source term, we have 
\begin{align}
    Z[J]=\int_{\C_{\P\T}}{\cal D}[\Phi]\exp\left(-\frac{1}{2}\int \d^d x\,\left[\partial_\mu\Phi\partial^\mu\Phi
    +\mu^2\Phi^2-(m^2/\phi_0^2)\Phi^4\right]-i\int \d^dx\, J \Phi\right)~,
\end{align}
where we again assume that $J$ is real.
Taking the complex conjugate of the above equation, one obtains

\begin{align}
    (Z[J])^*=\int_{\C^*_{\P\T}}{\cal D}[\Phi^*]\exp\left(-\frac{1}{2}\int \d^d x\,\left[\partial_\mu\Phi^*\partial^\mu\Phi^*
    +\mu^2\Phi^{*2}-(m^2/\phi_0^2)\Phi^{*4}\right]+i\int \d^dx\, J \Phi^*\right)~,
\end{align}
where $\C^*_{\P\T}$ is obtained from $\C_{\P\T}$ by taking complex conjugate of all its elements. The change of functional variable $\Phi^*\rightarrow -\Phi^*$, together with the condition~\eqref{eq:cond-contour}, finally lead to 
\begin{align}
    (Z[J])^*=Z[J]~.
\end{align}
Of course, the reality of the partition function is a consequence of its $\P\T$-invariance. Similarly, for the one-point function
\bea
\varphi[J]&\equiv& \langle \Phi\rangle\\
&\equiv& \frac{1}{Z[J]}\int_{\C_{\P\T}}{\cal D}[\Phi] 
\, \Phi\,\exp\left(-\frac{1}{2}\int \d^d x\,\left[\partial_\mu\Phi\partial^\mu\Phi+\mu^2\Phi^2-(m^2/\phi_0^2)\Phi^4\right]-i\int \d^dx\, J \Phi\right)~,\nonumber
\eea
we have $(\varphi[J])^*=-\varphi[J]$ so that the one-point function is purely imaginary. Again, one can extend this argument to arbitrary $n$-point functions to show that any $2N$-point correlation function is real and any $2N+1$-point correlation function is imaginary. The 1PI effective action, 
\be
\Gamma[\varphi]=-\ln Z[J[\varphi]]-i\int \d^d x\, \varphi J[\varphi]~,
\ee
is thus real. Using the above 1PI effective action, one can still carry out a derivation of the exact functional renormalisation equation, ending up with the same equation~\eqref{originalERG}.

\subsection{One-loop effective potential (\texorpdfstring{$d=4$}{TEXT})}

Given the above properties, the one-loop expression (\ref{Ueff1exp-4d}) can be used for $\varepsilon=2$ to give 
\bea
U_{\rm eff}^{(1)}(\varphi)&=&\frac{1}{2}\mu^2\varphi^2-\frac{\lambda}{24}\varphi^4
-\frac{\hbar}{64\pi^2}\Lambda^2\left(\mu^2-\lambda\varphi^2/2\right)\nn
&&-\frac{\hbar}{64\pi^2}\left(\mu^2-\lambda\varphi^2/2\right)^2
\left[\ln\left(\frac{\Lambda^2}{m^2}\right)-\ln\left(\frac{\mu^2-\lambda\varphi^2/2}{m^2}\right)\right]~,
\eea
where $\lambda\equiv 12m^2/\phi_0^2$.
The logarithmic and quadratic divergences can be absorbed by the introduction of the 
renormalised parameters 
\begin{subequations}
\begin{align}
   \mu_{\rm R}^2=&\mu^2-\frac{\hbar\lambda\Lambda^2}{64\pi^2}
+\frac{\hbar\lambda \mu^2}{16\pi^2}\ln\left(\frac{\Lambda}{m}\right)\\
\label{eq:lambdaR}
\lambda_{\rm R}=&\lambda+\frac{3\hbar\lambda^2}{16\pi^2}\ln\left(\frac{\Lambda}{m}\right)~, 
\end{align}
\end{subequations}
and the renormalised one-loop effective potential is, after ignoring the $\mathcal{O}(\hbar^2)$ and divergent but $\varphi$-independent terms,
\be\label{UR1e=2}
U_{\rm R}^{(1)}(\varphi)=\frac{1}{2}\mu_{\rm R}^2\varphi^2-\frac{\lambda_{\rm R}}{24}\varphi^4
+\frac{\hbar}{64\pi^2}\left(\mu_{\rm R}^2-\lambda_{\rm R}\varphi^2/2\right)^2\ln\left(\frac{\mu_{\rm R}^2-\lambda_{\rm R}\varphi^2/2}{m^2}\right)~.
\ee
Note that, since $\varphi$ is purely imaginary, the argument of the logarithm is always positive for $\mu_{\rm R}^2>0$,
and the effective potential (\ref{UR1e=2}) is always real. 
The possibility to define quantum corrections for the potential $-\phi^4$ is consistent with the conjecture that the effective theory 
does have a ground state at $\varphi=0$.

Also, quartic and logarithmic corrections come with the opposite sign
compared to the usual $+\phi^4$ theory. As a consequence, 
the interaction is asymptotically free; for a fixed renormalised coupling $\lambda_{\rm R}$, the bare coupling can be written as
\bea
\lambda&=&\lambda_{\rm R}-\frac{3\hbar\lambda^2}{16\pi^2}\ln\left(\frac{\Lambda}{m}\right)\\
&=&\lambda_{\rm R}-\frac{3\hbar\lambda_{\rm R}^2}{16\pi^2}\ln\left(\frac{\Lambda}{m}\right)+{\cal O}(\hbar^2)\nn
&=&\frac{\lambda_{\rm R}}{1+\frac{3\hbar \lambda_{\rm R}}{16\pi^2}\ln \left(\frac{\Lambda}{m}\right)}+{\cal O}(\hbar^2)~,\nonumber
\eea    
which, based on this one-loop result, leads to asymptotic freedom. Alternatively, the one-loop beta-function of the model is negative
\be
\beta^{(1)}\equiv\Lambda\partial_\Lambda\lambda=-\frac{3\hbar\lambda^2}{16\pi^2}~<0~,
\ee
and should be considered with the boundary condition $\lambda(m)=\lambda_{\rm R}$.

\subsection{One-loop effective potential (\texorpdfstring{$d=2$}{TEXT})}

Substituting $\varepsilon=2$ into Eq.~\eqref{eq:Ueff-2d}, we obtain
\bea
\label{eq:Ueff-2d-epsion2}
U_{\rm eff}^{(1)}(\varphi)=\frac{1}{2}\mu^2\varphi^2-\frac{\lambda}{24}\varphi^4+\frac{\hbar}{8\pi}\Big(\mu^2-\lambda \varphi^2/2\Big)\left[\ln\left(\frac{\Lambda^2}{m^2}\right)-\ln\left(\frac{\mu^2-\lambda\varphi^2/2}{m^2}\right)\right]\,,
\eea
where still $\lambda=12m^2/\phi_0^2$.
The logarithmically divergent quadratic term can be absorbed in a redefinition of $\mu^2$ through the renormalised mass squared
\be
\mu_{\rm R}^2=\mu^2+\frac{\hbar\,\lambda}{8\pi}\ln\left(\frac{\Lambda^2}{m^2}\right)~,
\ee
and the remaining terms are either finite or $\varphi$-independent. Ignoring the $\mathcal{O}(\hbar^2)$ and divergent but $\varphi$-independent terms, the renormalised effective potential reads
\begin{align}
\label{eq:Ueff-2d-epsion2-ren}
U_{\rm R}^{(1)}(\varphi)=\frac{1}{2}\mu^2_{\rm R}\varphi^2-\frac{\lambda}{24}\varphi^4
-\frac{\hbar}{8\pi}\left(\mu_{\rm R}^2-\lambda\varphi^2/2\right)\ln\left(\frac{\mu_{\rm R}^2-\lambda\varphi^2/2}{m^2}\right)\,.    
\end{align}

\section{Conclusion}
\label{sec:con}

$\P\T$-symmetric theories may open a new window to phenomenological model building for new physics. However, compared to quantum mechanics, quantum field theory is much more complicated since in the latter there are an infinite number of degrees of freedom. One particular issue that is absent from quantum-mechanical models but appears in quantum field theory is renormalisation. Therefore, when extending $\P\T$-symmetric quantum-mechanical models to quantum field theory, renormalisability has to be taken into account. In this paper, we have studied the renormalisation of a $\P\T$-symmetric scalar field theory with the bare potential given by Eq.~\eqref{Ubare}. This theory is a direct generalisation of the well-studied quantum-mechanical models given by the Hamiltonian~\eqref{eq:H}.

In contrast to Refs.~\cite{Bender:2018pbv,Felski:2021evi,Branchina:2021czr} where the renormalisation of the theory~\eqref{Ubare} is studied with the $\varepsilon$-expansion, our study is based on the Wilsonian approach and more specifically, the Wetterich equation for the running effective action. In this approach, the renormalisation of the theory~\eqref{Ubare} can be studied without doing the $\varepsilon$-expansion. We first carry out our analysis for regions of $\varepsilon$, e.g. $\epsilon\in[0,1/3]$ or $\epsilon=1$, in which the scalar field can be kept real in the Euclidean path integral. We have solved the Wetterich equation in the local potential approximation either in the UV regime or at the one-loop order. We obtained the scale-dependent one-loop effective potentials for arbitrary $\varepsilon$, Eq.~\eqref{eq:Uk-one-loop-4d} ($d=4$) and~\eqref{eq:Uk-one-loop-2d} ($d=2$), and their IR limits, Eqs.~\eqref{Ueff1exp-4d} and~\eqref{eq:Ueff-2d}.  We found that for a generic non-integer $\varepsilon$, one-loop corrections generate new interactions with divergent coefficients that cannot be absorbed by the bare parameters. Therefore at the one-loop level, the theory is renormalisable only for integer values of $\varepsilon$. The aforementioned general formulae are then applied particularly for $\varepsilon=1$. Although for $\varepsilon=2$, a deformation for the integration contour of the Euclidean path integral is necessary to ensure the convergence of the latter, we argue that the general formulae can still be applied for $\varepsilon=2$. One-loop beta functions for the coupling associated with the interaction $i\phi^3$ and $-\phi^4$ are then computed. It is confirmed that the $-\phi^4$ theory has asymptotic freedom in four-dimensional spacetime. We have also shown that a consequence of $\P\T$ symmetry is that all the odd-point correlation functions are imaginary and all the even-point correlation functions, including the partition function itself, are real. 

\section*{Acknowledgements}
This work is supported by the UK Engineering and Physical Sciences Research Council (grant EP/V002821/1), and the Science and Technology
Facilities Council (grant STFC-ST/T000759/1).

\bibliographystyle{utphys}
\bibliography{Ref}{}

\providecommand{\href}[2]{#2}\begingroup\raggedright\begin{thebibliography}{10}

\bibitem{Bender:1998ke}
C.~M. Bender and S.~Boettcher, ``{Real spectra in nonHermitian Hamiltonians
  having PT symmetry},''
  \href{http://dx.doi.org/10.1103/PhysRevLett.80.5243}{{\em Phys. Rev. Lett.}
  {\bfseries 80} (1998) 5243--5246},
  \href{http://arxiv.org/abs/physics/9712001}{{\ttfamily
  arXiv:physics/9712001}}.

\bibitem{Bender:2002vv}
C.~M. Bender, D.~C. Brody, and H.~F. Jones, ``{Complex extension of quantum
  mechanics},'' \href{http://dx.doi.org/10.1103/PhysRevLett.89.270401}{{\em
  Phys. Rev. Lett.} {\bfseries 89} (2002) 270401},
  \href{http://arxiv.org/abs/quant-ph/0208076}{{\ttfamily
  arXiv:quant-ph/0208076}}. [Erratum: Phys.Rev.Lett. 92, 119902 (2004)].

\bibitem{Bender:2007nj}
C.~M. Bender, ``{Making sense of non-Hermitian Hamiltonians},''
  \href{http://dx.doi.org/10.1088/0034-4885/70/6/R03}{{\em Rept. Prog. Phys.}
  {\bfseries 70} (2007) 947},
  \href{http://arxiv.org/abs/hep-th/0703096}{{\ttfamily arXiv:hep-th/0703096}}.

\bibitem{christodoulides2018parity}
D.~Christodoulides, J.~Yang, {\em et~al.}, {\em {Parity-time symmetry and its
  applications}}, vol.~280.
\newblock Springer, Singapore, 2018.

\bibitem{Bender:2019cwm}
C.~M. Bender, P.~E. Dorey, C.~Dunning, A.~Fring, D.~W. Hook, H.~F. Jones,
  S.~Kuzhel, G.~L\'evai, and R.~Tateo,
  \href{http://dx.doi.org/10.1142/q0178}{{\em {PT Symmetry in Quantum and
  Classical Physics}}}.
\newblock World Scientific, Singapore, 2019.

\bibitem{longhi2018parity}
S.~Longhi, ``{Parity-time symmetry meets photonics: A new twist in
  non-Hermitian optics},'' {\em EPL} {\bfseries 120} no.~6, (2018) 64001,
  \href{http://arxiv.org/abs/1802.05025}{{\ttfamily arXiv:1802.05025}}.

\bibitem{el2018non}
R.~El-Ganainy, K.~G. Makris, M.~Khajavikhan, Z.~H. Musslimani, S.~Rotter, and
  D.~N. Christodoulides, ``{Non-Hermitian physics and PT symmetry},'' {\em
  Nature Physics} {\bfseries 14} no.~1, (2018) 11--19.

\bibitem{Ashida:2020dkc}
Y.~Ashida, Z.~Gong, and M.~Ueda, ``{Non-Hermitian physics},''
  \href{http://dx.doi.org/10.1080/00018732.2021.1876991}{{\em Adv. Phys.}
  {\bfseries 69} no.~3, (2021) 249--435},
  \href{http://arxiv.org/abs/2006.01837}{{\ttfamily arXiv:2006.01837
  [cond-mat.mes-hall]}}.

\bibitem{Jones-Smith:2009qeu}
K.~Jones-Smith and H.~Mathur, ``{Relativistic Non-Hermitian Quantum
  Mechanics},'' \href{http://dx.doi.org/10.1103/PhysRevD.89.125014}{{\em Phys.
  Rev. D} {\bfseries 89} no.~12, (2014) 125014},
  \href{http://arxiv.org/abs/0908.4257}{{\ttfamily arXiv:0908.4257 [hep-th]}}.

\bibitem{Ohlsson:2015xsa}
T.~Ohlsson, ``{Non-Hermitian neutrino oscillations in matter with PT symmetric
  Hamiltonians},'' \href{http://dx.doi.org/10.1209/0295-5075/113/61001}{{\em
  EPL} {\bfseries 113} no.~6, (2016) 61001},
  \href{http://arxiv.org/abs/1509.06452}{{\ttfamily arXiv:1509.06452
  [hep-ph]}}.

\bibitem{Alexandre:2015kra}
J.~Alexandre, C.~M. Bender, and P.~Millington, ``{Non-Hermitian extension of
  gauge theories and implications for neutrino physics},''
  \href{http://dx.doi.org/10.1007/JHEP11(2015)111}{{\em JHEP} {\bfseries 11}
  (2015) 111}, \href{http://arxiv.org/abs/1509.01203}{{\ttfamily
  arXiv:1509.01203 [hep-th]}}.

\bibitem{Rodionov:2017dqt}
V.~N. Rodionov and A.~M. Mandel, ``{An upper limit on fermion mass spectrum in
  non-Hermitian models and its implications for studying of dark matter},''
  \href{http://arxiv.org/abs/1708.08394}{{\ttfamily arXiv:1708.08394
  [hep-ph]}}.

\bibitem{Korchin:2016rsf}
A.~Y. Korchin and V.~A. Kovalchuk, ``{Decay of the Higgs boson to $\tau^-
  \tau^+$ and non-Hermiticy of the Yukawa interaction},''
  \href{http://dx.doi.org/10.1103/PhysRevD.94.076003}{{\em Phys. Rev. D}
  {\bfseries 94} no.~7, (2016) 076003},
  \href{http://arxiv.org/abs/1607.02827}{{\ttfamily arXiv:1607.02827
  [hep-ph]}}.

\bibitem{Raval:2018kqg}
H.~Raval and B.~P. Mandal, ``{Deconfinement to Confinement as PT phase
  transition},'' \href{http://dx.doi.org/10.1016/j.nuclphysb.2019.114699}{{\em
  Nucl. Phys.} {\bfseries B} (2019) 114699},
  \href{http://arxiv.org/abs/1805.02510}{{\ttfamily arXiv:1805.02510
  [hep-th]}}.

\bibitem{Alexandre:2017foi}
J.~Alexandre, P.~Millington, and D.~Seynaeve, ``{Symmetries and conservation
  laws in non-Hermitian field theories},''
  \href{http://dx.doi.org/10.1103/PhysRevD.96.065027}{{\em Phys. Rev. D}
  {\bfseries 96} no.~6, (2017) 065027},
  \href{http://arxiv.org/abs/1707.01057}{{\ttfamily arXiv:1707.01057
  [hep-th]}}.

\bibitem{Alexandre:2018uol}
J.~Alexandre, J.~Ellis, P.~Millington, and D.~Seynaeve, ``{Spontaneous symmetry
  breaking and the Goldstone theorem in non-Hermitian field theories},''
  \href{http://dx.doi.org/10.1103/PhysRevD.98.045001}{{\em Phys. Rev. D}
  {\bfseries 98} (2018) 045001},
  \href{http://arxiv.org/abs/1805.06380}{{\ttfamily arXiv:1805.06380
  [hep-th]}}.

\bibitem{Mannheim:2018dur}
P.~D. Mannheim, ``{Goldstone bosons and the Englert-Brout-Higgs mechanism in
  non-Hermitian theories},''
  \href{http://dx.doi.org/10.1103/PhysRevD.99.045006}{{\em Phys. Rev. D}
  {\bfseries 99} no.~4, (2019) 045006},
  \href{http://arxiv.org/abs/1808.00437}{{\ttfamily arXiv:1808.00437
  [hep-th]}}.

\bibitem{Alexandre:2018xyy}
J.~Alexandre, J.~Ellis, P.~Millington, and D.~Seynaeve, ``{Gauge invariance and
  the Englert-Brout-Higgs mechanism in non-Hermitian field theories},''
  \href{http://dx.doi.org/10.1103/PhysRevD.99.075024}{{\em Phys. Rev. D}
  {\bfseries 99} no.~7, (2019) 075024},
  \href{http://arxiv.org/abs/1808.00944}{{\ttfamily arXiv:1808.00944
  [hep-th]}}.

\bibitem{Fring:2019hue}
A.~Fring and T.~Taira, ``{Goldstone bosons in different PT-regimes of
  non-Hermitian scalar quantum field theories},''
  \href{http://dx.doi.org/10.1016/j.nuclphysb.2019.114834}{{\em Nucl. Phys. B}
  {\bfseries 950} (2020) 114834},
  \href{http://arxiv.org/abs/1906.05738}{{\ttfamily arXiv:1906.05738
  [hep-th]}}.

\bibitem{Fring:2019xgw}
A.~Fring and T.~Taira, ``{Pseudo-Hermitian approach to
  Goldstone\textquoteright{}s theorem in non-Abelian non-Hermitian quantum
  field theories},'' \href{http://dx.doi.org/10.1103/PhysRevD.101.045014}{{\em
  Phys. Rev. D} {\bfseries 101} no.~4, (2020) 045014},
  \href{http://arxiv.org/abs/1911.01405}{{\ttfamily arXiv:1911.01405
  [hep-th]}}.

\bibitem{Alexandre:2019jdb}
J.~Alexandre, J.~Ellis, P.~Millington, and D.~Seynaeve, ``{Spontaneously
  Breaking Non-Abelian Gauge Symmetry in Non-Hermitian Field Theories},''
  \href{http://dx.doi.org/10.1103/PhysRevD.101.035008}{{\em Phys. Rev. D}
  {\bfseries 101} no.~3, (2020) 035008},
  \href{http://arxiv.org/abs/1910.03985}{{\ttfamily arXiv:1910.03985
  [hep-th]}}.

\bibitem{Fring:2020bvr}
A.~Fring and T.~Taira, ``{Massive gauge particles versus Goldstone bosons in
  non-Hermitian non-Abelian gauge theory},''
  \href{http://dx.doi.org/10.1140/epjp/s13360-022-02889-z}{{\em Eur. Phys. J.
  Plus} {\bfseries 137} no.~6, (2022) 716},
  \href{http://arxiv.org/abs/2004.00723}{{\ttfamily arXiv:2004.00723
  [hep-th]}}.

\bibitem{Alexandre:2020bet}
J.~Alexandre and N.~E. Mavromatos, ``{On the consistency of a non-Hermitian
  Yukawa interaction},''
  \href{http://dx.doi.org/10.1016/j.physletb.2020.135562}{{\em Phys. Lett. B}
  {\bfseries 807} (2020) 135562},
  \href{http://arxiv.org/abs/2004.03699}{{\ttfamily arXiv:2004.03699
  [hep-ph]}}.

\bibitem{Alexandre:2020tba}
J.~Alexandre, N.~E. Mavromatos, and A.~Soto, ``{Dynamical Majorana neutrino
  masses and axions I},''
  \href{http://dx.doi.org/10.1016/j.nuclphysb.2020.115212}{{\em Nucl. Phys. B}
  {\bfseries 961} (2020) 115212},
  \href{http://arxiv.org/abs/2004.04611}{{\ttfamily arXiv:2004.04611
  [hep-ph]}}.

\bibitem{Mavromatos:2020hfy}
N.~E. Mavromatos and A.~Soto, ``{Dynamical Majorana neutrino masses and axions
  II: Inclusion of anomaly terms and axial background},''
  \href{http://dx.doi.org/10.1016/j.nuclphysb.2020.115275}{{\em Nucl. Phys. B}
  {\bfseries 962} (2021) 115275},
  \href{http://arxiv.org/abs/2006.13616}{{\ttfamily arXiv:2006.13616
  [hep-ph]}}.

\bibitem{Mavromatos:2021hpe}
N.~E. Mavromatos, S.~Sarkar, and A.~Soto, ``{PT symmetric fermionic field
  theories with axions: Renormalization and dynamical mass generation},''
  \href{http://dx.doi.org/10.1103/PhysRevD.106.015009}{{\em Phys. Rev. D}
  {\bfseries 106} no.~1, (2022) 015009},
  \href{http://arxiv.org/abs/2111.05131}{{\ttfamily arXiv:2111.05131
  [hep-th]}}.

\bibitem{Alexandre:2020gah}
J.~Alexandre, J.~Ellis, and P.~Millington, ``{Discrete spacetime symmetries and
  particle mixing in non-Hermitian scalar quantum field theories},''
  \href{http://dx.doi.org/10.1103/PhysRevD.102.125030}{{\em Phys. Rev. D}
  {\bfseries 102} no.~12, (2020) 125030},
  \href{http://arxiv.org/abs/2006.06656}{{\ttfamily arXiv:2006.06656
  [hep-th]}}.

\bibitem{Alexandre:2022uns}
J.~Alexandre, J.~Ellis, and P.~Millington, ``{Discrete spacetime symmetries,
  second quantization, and inner products in a non-Hermitian Dirac fermionic
  field theory},'' \href{http://dx.doi.org/10.1103/PhysRevD.106.065003}{{\em
  Phys. Rev. D} {\bfseries 106} no.~6, (2022) 065003},
  \href{http://arxiv.org/abs/2201.11061}{{\ttfamily arXiv:2201.11061
  [hep-th]}}.

\bibitem{Beygi:2019qab}
A.~Beygi, S.~P. Klevansky, and C.~M. Bender, ``{Relativistic $PT$-symmetric
  fermionic theories in 1+1 and 3+1 dimensions},''
  \href{http://dx.doi.org/10.1103/PhysRevA.99.062117}{{\em Phys. Rev. A}
  {\bfseries 99} no.~6, (2019) 062117},
  \href{http://arxiv.org/abs/1904.00878}{{\ttfamily arXiv:1904.00878
  [math-ph]}}.

\bibitem{Alexandre:2020wki}
J.~Alexandre, J.~Ellis, and P.~Millington, ``{$\mathcal{PT}$ -symmetric
  non-Hermitian quantum field theories with supersymmetry},''
  \href{http://dx.doi.org/10.1103/PhysRevD.101.085015}{{\em Phys. Rev. D}
  {\bfseries 101} no.~8, (2020) 085015},
  \href{http://arxiv.org/abs/2001.11996}{{\ttfamily arXiv:2001.11996
  [hep-th]}}.

\bibitem{Felski:2020vrm}
A.~Felski, A.~Beygi, and S.~P. Klevansky, ``{Non-Hermitian extension of the
  Nambu\textendash{}Jona-Lasinio model in 3+1 and 1+1 dimensions},''
  \href{http://dx.doi.org/10.1103/PhysRevD.101.116001}{{\em Phys. Rev. D}
  {\bfseries 101} no.~11, (2020) 116001},
  \href{http://arxiv.org/abs/2004.04011}{{\ttfamily arXiv:2004.04011
  [hep-ph]}}.

\bibitem{Fring:2020xpi}
A.~Fring and T.~Taira, ``{'t Hooft-Polyakov monopoles in non-Hermitian quantum
  field theory},'' \href{http://dx.doi.org/10.1016/j.physletb.2020.135583}{{\em
  Phys. Lett. B} {\bfseries 807} (2020) 135583},
  \href{http://arxiv.org/abs/2006.02718}{{\ttfamily arXiv:2006.02718
  [hep-th]}}.

\bibitem{Chernodub:2020cew}
M.~N. Chernodub, A.~Cortijo, and M.~Ruggieri, ``{Spontaneous non-Hermiticity in
  the Nambu\textendash{}Jona-Lasinio model},''
  \href{http://dx.doi.org/10.1103/PhysRevD.104.056023}{{\em Phys. Rev. D}
  {\bfseries 104} no.~5, (2021) 056023},
  \href{http://arxiv.org/abs/2008.11629}{{\ttfamily arXiv:2008.11629
  [hep-th]}}.

\bibitem{Fring:2021zci}
A.~Fring and T.~Taira, ``{Non-Hermitian gauge field theories and BPS limits},''
  \href{http://dx.doi.org/10.1088/1742-6596/2038/1/012010}{{\em J. Phys. Conf.
  Ser.} {\bfseries 2038} (2021) 012010},
  \href{http://arxiv.org/abs/2103.13519}{{\ttfamily arXiv:2103.13519
  [hep-th]}}.

\bibitem{Felski:2021bdg}
A.~Felski and S.~P. Klevansky, ``{Fermion and meson mass generation in
  non-Hermitian Nambu\textendash{}Jona-Lasinio models},''
  \href{http://dx.doi.org/10.1103/PhysRevD.103.056007}{{\em Phys. Rev. D}
  {\bfseries 103} no.~5, (2021) 056007},
  \href{http://arxiv.org/abs/2102.01491}{{\ttfamily arXiv:2102.01491
  [hep-ph]}}.

\bibitem{Felski:2021evi}
A.~Felski, C.~M. Bender, S.~P. Klevansky, and S.~Sarkar, ``{Towards
  perturbative renormalization of
  \ensuremath{\phi}2(i\ensuremath{\phi})\ensuremath{\epsilon} quantum field
  theory},'' \href{http://dx.doi.org/10.1103/PhysRevD.104.085011}{{\em Phys.
  Rev. D} {\bfseries 104} no.~8, (2021) 085011},
  \href{http://arxiv.org/abs/2103.07577}{{\ttfamily arXiv:2103.07577
  [hep-th]}}.

\bibitem{Khunjua:2021fus}
T.~G. Khunjua, K.~G. Klimenko, and R.~N. Zhokhov, ``{Spontaneous
  non-Hermiticity in the (2+1)-dimensional Gross-Neveu model},''
  \href{http://dx.doi.org/10.1103/PhysRevD.105.025014}{{\em Phys. Rev. D}
  {\bfseries 105} no.~2, (2022) 025014},
  \href{http://arxiv.org/abs/2112.13012}{{\ttfamily arXiv:2112.13012
  [hep-th]}}.

\bibitem{Mavromatos:2022heh}
N.~E. Mavromatos, S.~Sarkar, and A.~Soto, ``{Schwinger-Dyson equations and mass
  generation for an axion theory with a PT symmetric Yukawa fermion
  interaction},'' \href{http://arxiv.org/abs/2208.12436}{{\ttfamily
  arXiv:2208.12436 [hep-ph]}}.

\bibitem{Felski:2022dsx}
A.~Felski, A.~Beygi, and S.~P. Klevansky, ``{Thermodynamic properties of
  non-Hermitian Nambu--Jona-Lasinio models},''
  \href{http://arxiv.org/abs/2210.15503}{{\ttfamily arXiv:2210.15503
  [hep-ph]}}.

\bibitem{Naon:2022xvl}
C.~M. Na\'on and F.~A. Schaposnik, ``{Path-integral Bosonization of $d=2$ PT
  symmetric models},'' \href{http://arxiv.org/abs/2211.02978}{{\ttfamily
  arXiv:2211.02978 [hep-th]}}.

\bibitem{Gubaeva:2022feb}
M.~M. Gubaeva, T.~G. Khunjua, K.~G. Klimenko, and R.~N. Zhokhov, ``{Spontaneous
  non-Hermiticity in the (2+1)-dimensional Thirring model},''
  \href{http://dx.doi.org/10.1103/PhysRevD.106.125010}{{\em Phys. Rev. D}
  {\bfseries 106} no.~12, (2022) 125010},
  \href{http://arxiv.org/abs/2212.01062}{{\ttfamily arXiv:2212.01062
  [hep-th]}}.

\bibitem{Dorey:2001uw}
P.~Dorey, C.~Dunning, and R.~Tateo, ``{Spectral equivalences, Bethe Ansatz
  equations, and reality properties in PT-symmetric quantum mechanics},''
  \href{http://dx.doi.org/10.1088/0305-4470/34/28/305}{{\em J. Phys. A}
  {\bfseries 34} (2001) 5679--5704},
  \href{http://arxiv.org/abs/hep-th/0103051}{{\ttfamily arXiv:hep-th/0103051}}.

\bibitem{Jones:2006qs}
H.~F. Jones and J.~Mateo, ``{An Equivalent Hermitian Hamiltonian for the
  non-Hermitian $-x^4$ potential},''
  \href{http://dx.doi.org/10.1103/PhysRevD.73.085002}{{\em Phys. Rev. D}
  {\bfseries 73} (2006) 085002},
  \href{http://arxiv.org/abs/quant-ph/0601188}{{\ttfamily
  arXiv:quant-ph/0601188}}.

\bibitem{Bender:2006wt}
C.~M. Bender, D.~C. Brody, J.-H. Chen, H.~F. Jones, K.~A. Milton, and M.~C.
  Ogilvie, ``{Equivalence of a Complex PT-Symmetric Quartic Hamiltonian and a
  Hermitian Quartic Hamiltonian with an Anomaly},''
  \href{http://dx.doi.org/10.1103/PhysRevD.74.025016}{{\em Phys. Rev. D}
  {\bfseries 74} (2006) 025016},
  \href{http://arxiv.org/abs/hep-th/0605066}{{\ttfamily arXiv:hep-th/0605066}}.

\bibitem{Bender:2018pbv}
C.~M. Bender, N.~Hassanpour, S.~P. Klevansky, and S.~Sarkar, ``{PT-symmetric
  quantum field theory in D dimensions},''
  \href{http://dx.doi.org/10.1103/PhysRevD.98.125003}{{\em Phys. Rev. D}
  {\bfseries 98} no.~12, (2018) 125003},
  \href{http://arxiv.org/abs/1810.12479}{{\ttfamily arXiv:1810.12479
  [hep-th]}}.

\bibitem{Bender:2021fxa}
C.~M. Bender, A.~Felski, S.~P. Klevansky, and S.~Sarkar, ``{PT Symmetry and
  Renormalisation in Quantum Field Theory},''
  \href{http://arxiv.org/abs/2103.14864}{{\ttfamily arXiv:2103.14864
  [hep-th]}}.

\bibitem{Ai:2022csx}
W.-Y. Ai, C.~M. Bender, and S.~Sarkar, ``{PT-symmetric -g\ensuremath{\varphi}4
  theory},'' \href{http://dx.doi.org/10.1103/PhysRevD.106.125016}{{\em Phys.
  Rev. D} {\bfseries 106} no.~12, (2022) 125016},
  \href{http://arxiv.org/abs/2209.07897}{{\ttfamily arXiv:2209.07897
  [hep-th]}}.

\bibitem{Branchina:2021czr}
V.~Branchina, A.~Chiavetta, and F.~Contino, ``{Study of the non-Hermitian
  PT-symmetric g\ensuremath{\phi}2(i\ensuremath{\phi})\ensuremath{\varepsilon}
  theory: Analysis of all orders in \ensuremath{\varepsilon} and
  resummations},'' \href{http://dx.doi.org/10.1103/PhysRevD.104.085010}{{\em
  Phys. Rev. D} {\bfseries 104} no.~8, (2021) 085010},
  \href{http://arxiv.org/abs/2104.12702}{{\ttfamily arXiv:2104.12702
  [hep-th]}}.

\bibitem{Bender:1987dn}
C.~M. Bender, K.~A. Milton, M.~Moshe, S.~S. Pinsky, and L.~M. Simmons,
  ``{LOGARITHMIC APPROXIMATIONS TO POLYNOMIAL LAGRANGIANS},''
  \href{http://dx.doi.org/10.1103/PhysRevLett.58.2615}{{\em Phys. Rev. Lett.}
  {\bfseries 58} (1987) 2615--2618}.

\bibitem{Bender:1988rq}
C.~M. Bender, K.~A. Milton, M.~Moshe, S.~S. Pinsky, and L.~M. Simmons, Jr.,
  ``{Novel Perturbative Scheme in Quantum Field Theory},''
  \href{http://dx.doi.org/10.1103/PhysRevD.37.1472}{{\em Phys. Rev. D}
  {\bfseries 37} (1988) 1472}.

\bibitem{Wegner:1972ih}
F.~J. Wegner and A.~Houghton, ``{Renormalization group equation for critical
  phenomena},'' \href{http://dx.doi.org/10.1103/PhysRevA.8.401}{{\em Phys. Rev.
  A} {\bfseries 8} (1973) 401--412}.

\bibitem{Polchinski:1983gv}
J.~Polchinski, ``{Renormalization and Effective Lagrangians},''
  \href{http://dx.doi.org/10.1016/0550-3213(84)90287-6}{{\em Nucl. Phys. B}
  {\bfseries 231} (1984) 269--295}.

\bibitem{Wetterich:2001kra}
C.~Wetterich, ``{Effective average action in statistical physics and quantum
  field theory},'' \href{http://dx.doi.org/10.1142/S0217751X01004591}{{\em Int.
  J. Mod. Phys. A} {\bfseries 16} (2001) 1951--1982},
  \href{http://arxiv.org/abs/hep-ph/0101178}{{\ttfamily arXiv:hep-ph/0101178}}.

\bibitem{pham1983vanishing}
F.~Pham, ``Vanishing homologies and the n variables saddlepoint method,'' in
  {\em Proc. Symp. Pure Math.}, vol.~40, pp.~310--333.
\newblock 1983.

\bibitem{berry1991hyperasymptotics}
M.~V. Berry and C.~J. Howls, ``Hyperasymptotics for integrals with saddles,''
  {\em Proceedings of the Royal Society of London. Series A: Mathematical and
  Physical Sciences} {\bfseries 434} no.~1892, (1991) 657--675.

\bibitem{Witten:2010cx}
E.~Witten, ``{Analytic Continuation Of Chern-Simons Theory},'' {\em AMS/IP
  Stud. Adv. Math.} {\bfseries 50} (2011) 347--446,
  \href{http://arxiv.org/abs/1001.2933}{{\ttfamily arXiv:1001.2933 [hep-th]}}.

\bibitem{Witten:2010zr}
E.~Witten, ``{A New Look At The Path Integral Of Quantum Mechanics},''
  \href{http://arxiv.org/abs/1009.6032}{{\ttfamily arXiv:1009.6032 [hep-th]}}.

\bibitem{Ai:2019fri}
W.-Y. Ai, B.~Garbrecht, and C.~Tamarit, ``{Functional methods for false vacuum
  decay in real time},'' \href{http://dx.doi.org/10.1007/JHEP12(2019)095}{{\em
  JHEP} {\bfseries 12} (2019) 095},
  \href{http://arxiv.org/abs/1905.04236}{{\ttfamily arXiv:1905.04236
  [hep-th]}}.

\bibitem{Jackiw:1974cv}
R.~Jackiw, ``{Functional evaluation of the effective potential},''
  \href{http://dx.doi.org/10.1103/PhysRevD.9.1686}{{\em Phys. Rev. D}
  {\bfseries 9} (1974) 1686}.

\bibitem{Litim:2001up}
D.~F. Litim, ``{Optimized renormalization group flows},''
  \href{http://dx.doi.org/10.1103/PhysRevD.64.105007}{{\em Phys. Rev. D}
  {\bfseries 64} (2001) 105007},
  \href{http://arxiv.org/abs/hep-th/0103195}{{\ttfamily arXiv:hep-th/0103195}}.

\bibitem{Bender:2018xpo}
C.~M. Bender and S.~Sarkar, ``{Asymptotic Analysis of the Local Potential
  Approximation to the Wetterich Equation},''
  \href{http://dx.doi.org/10.1088/1751-8121/aabf63}{{\em J. Phys. A} {\bfseries
  51} no.~22, (2018) 225202}, \href{http://arxiv.org/abs/1801.08106}{{\ttfamily
  arXiv:1801.08106 [hep-th]}}.

\end{thebibliography}\endgroup

\end{document}